\documentclass{ptapap}
\usepackage{amsmath}
\usepackage{amssymb}

\author{Agnieszka Janiuk}[CFT]
\affil[CFT]{Center for Theortical Physics, Polish Academy of Sciences, Al. Lotnikow 32/46, 02--668, Warsaw, Poland}

\title{Many faces of accretion in gamma ray bursts}

\begin{document}

\maketitle

\begin{abstract}
  Accretion powers relativistic jets in GRBs, similarly to other jet sources.
  Black holes that are at heart of long GRBs, are formed as the end product of stellar evolution. 
  At birth, some of the black holes must be very rapidly spinning, to be able to power the GRBS.
 In some cases, the black holes may be born without formation of a disk/jet engine, and then the star will collapse without an electromagnetic transient.
In this proceeding, we discuss the conditions for launching variable jets from the magnetized disk in an arrested state. 
We also discuss properties of collapsing massive stars as
progenitors of GRBs, and the conditions which must be satisfied for the star
in order for the collapsar to produce a bright gamma-ray transient.
We find that the black hole rotation is further affected by
self-gravity of the collapsing matter.
Finally, we comment on the properties of the accretion disk under extreme conditions of nuclear densities and temperatures, while it can contribute to the kilonova accompanying short GRBs.
\end{abstract}

\section{Introduction}

\subsection{Gamma Ray Bursts}

Massive stars ($M>8 M_{\odot}$) end their lives as core-collapse (CC) supernovae.
The subtypes of CC SNe are indicating the amount of hydrogen and helium-rich envelope that remains at their deaths.
Single stars are prone to the mass loss via stellar winds, but this mechanism cannot explain the diversity of all CC SNe.
Stellar winds are luminosity driven
and only stars with mass above 30 Solar mass can lose their H-rich envelope this way. Otherwise, the majority of stars that produce hydrogen- and helium-poor
stripped-envelope SNe are not that massive, while they are making about
37\% of all CC SNe \citep{2011MNRAS.412.1522S}. Therefore, other physical mechanisms, from nuclear instabilities to binary interactions, must contribute to the ejection of mass
from these stars prior to their explosions.

The remnant left by the collapsed star may either be a neutron star or a black hole. The latter provides a plausible mechanism of launching jets via extraction of black hole's rotational energy. There are observational evidences that choked jets are active in CCSNe that are not associated with GRBs \citep{Piran2019}.
The question remains however, why only about 1\% of CC SNe are accompanied by GRBs and what are the necessary conditions in the star to support both jet launching and breakout.
Recent numerical simulations show that only
a limited number of states in the progenitor’s parameter space supports relativistic jet launching, requiring both fast rotation and strong magnetic field \citep{Gottlieb2021}.

Another class of gamma ray bursts originate from the coalescence of binary
compact objects, where also the disk/jet engine is formed.
The electromagnetic radiation in gamma rays is accompanied by a gravitational wave signal from the binary merger, while the accretion disk uncollimated outflows may contribute to the subrelativistic neutron-rich material, which is responsible for blue and red kilonova \citep{LiPaczynski1998}.
It is still debated how exactly the short gamma-ray burst is launched after the merger.
However magnetic fields may have negligible effects on the evolution in the inspiral phase \citep{Giacomazzo2009}, the MHD instabilities and turbulence could significantly
affect the structure
and strength of the magnetic fields during and after the merger.
The 
fields during the merger can be amplified by several orders of magnitude up to
$\sim 10^{16}$ Gauss \citep{Palenzuela2015} and such extreme magnetic fields can drive the jets.
A metastable strongly
magnetised hypermassive neutron star (HMNS), which is more
massive than the maximum mass supported by the nuclear equation of state in the spherical and
uniformly rotating configuration, e.g., the proto-magnetar, is likely formed after the merger and then collapse to a black hole. Hence, a delay between the merger and GRB prompt emission is expected.
The neutron-rich materials, which are ejected during and after the merger
power kilonovae \citep{Metzger2019LRR}. The details of the kilonova spectra depend on the mass, velocity, and electron fraction distribution in the ejecta, that are launched during the merger and from the accretion disk.

\subsection{Numerical Relativity}

Two important measurements of the last decade, the first detection of the binary black hole merger by LIGO \citep{bbh}, and the first detection of multi-messenger signals associated with a binary neutron star merger \citep{bns,2017Swope,2017Fermi}, have made the numerical relativity methods essential for modern astrophysics.
Characterizing GW signals requires accurate template waveforms for the inspiral, merger, and ringdown phase of binary black hole coalescence. Producing these waveforms is a theoretical challenge both in General Relativity (GR) and in potential modifications of it. 
As the two-body problem has no analytical solutions in General Relativity, the numerical methods solving the accelerated motion of black hole binary were developed, based on a generalization of harmonic coordinates \citep{Pretorius2005},
or on  handling of the singular puncture conformal factor by an initial 'lapse'  \citep{Campanelli2006}, to avoid singularities during the evolution.
The subsequent usage of adaptive mesh refinement techniques provided adequate resolution for an accurate estimate of the energy emitted in gravitational waves.

Numerical simulations allow to study in detail the process of binary neutron star, or black hole-neutron star mergers.
The general relativistic magnetohydrodynamics and multi-dimensional, multi-scale calculations are essential to produce a self-consistent picture of jet and other ejecta accompanying the compact object coalescence. These ejecta give rise to the multi-messenger observations in broad-band electromagnetic spectra, and possibly neutrinos, and are sensitively dependent on the binary parameters constrained by the GW detections.

The general relativistic magneto-hydrodynamic code, \textit{HARM} \citep{Gammie_2003, Noble_et_all_2006}, is a finite volume, shock capturing scheme that solves 
the hyperbolic system of the partial differential equations of GR MHD. 
The plasma energy-momentum tensor, $T^{\mu\nu}$, is contributed by gas and electromagnetic field: $T^{\mu\nu}={T_{\left(m\right)}}^{\mu\nu}+{T_{\left(em\right)}}^{\mu\nu}$.

\begin{equation}
  \
{T_{\left(m\right)}}^{\mu\nu}= \rho h u^\mu u^\nu + p g^{\mu\nu};
\hspace{1cm}
{T_{\left(em\right)}}^{\mu\nu}=b^\kappa b_\kappa u^\mu u^\nu+\frac{1}{2} b^\kappa b_\kappa g^{\mu\nu} - b^\mu b^\nu.
\end{equation}
where $u^{\mu}$ is the four-velocity of gas, $u$ denotes internal energy density,  $b^{\mu}$ is magnetic four-vector, and
%$\enth$ 
$h$ is the fluid specific enthalpy.
The continuity and momentum conservation equations read:
\begin{equation}
\
(\rho u^{\mu})_{;\mu} = 0;
\hspace{1cm}
T^{\mu}_{\nu;\mu} = 0.
\end{equation}
They are brought in conservative form, by implementing a Harten, Lax, van Leer (HLL) solver to calculate numerically the corresponding fluxes.

In the current work, we use our personal version of this scheme, which was
extended to 3D and parallelized with hybrid MPI-OpenMP technique to enable efficient calculations. The microphysics of dense nuclear matter in the GRB engine is treated with specific tabulated EOS module, as introduced in
\cite{Janiuk2017}.
We also implement the methods developed in \citep{2018ApJ...868...68J}  to account for the change of Kerr metric due to the dynamically increasing black hole mass and its spin.

\section{GRB jet launching}

Relativistic jets that emit gamma rays, emerge during the catastrophic deaths of massive stars and binary neutron star mergers.
The physics of these systems is revealed by numerical magnetohydrodynamic simulations, often combined with semi-analytical calculations of radiation emission.
In this Section, we present some recent results of our simulation of jet launching from the rotating Kerr black hole, surrounded by the magnetically-arrested disk.

\begin{figure}
  \centering
  \begin{minipage}{0.45\textwidth}
    \includegraphics[width=\textwidth]{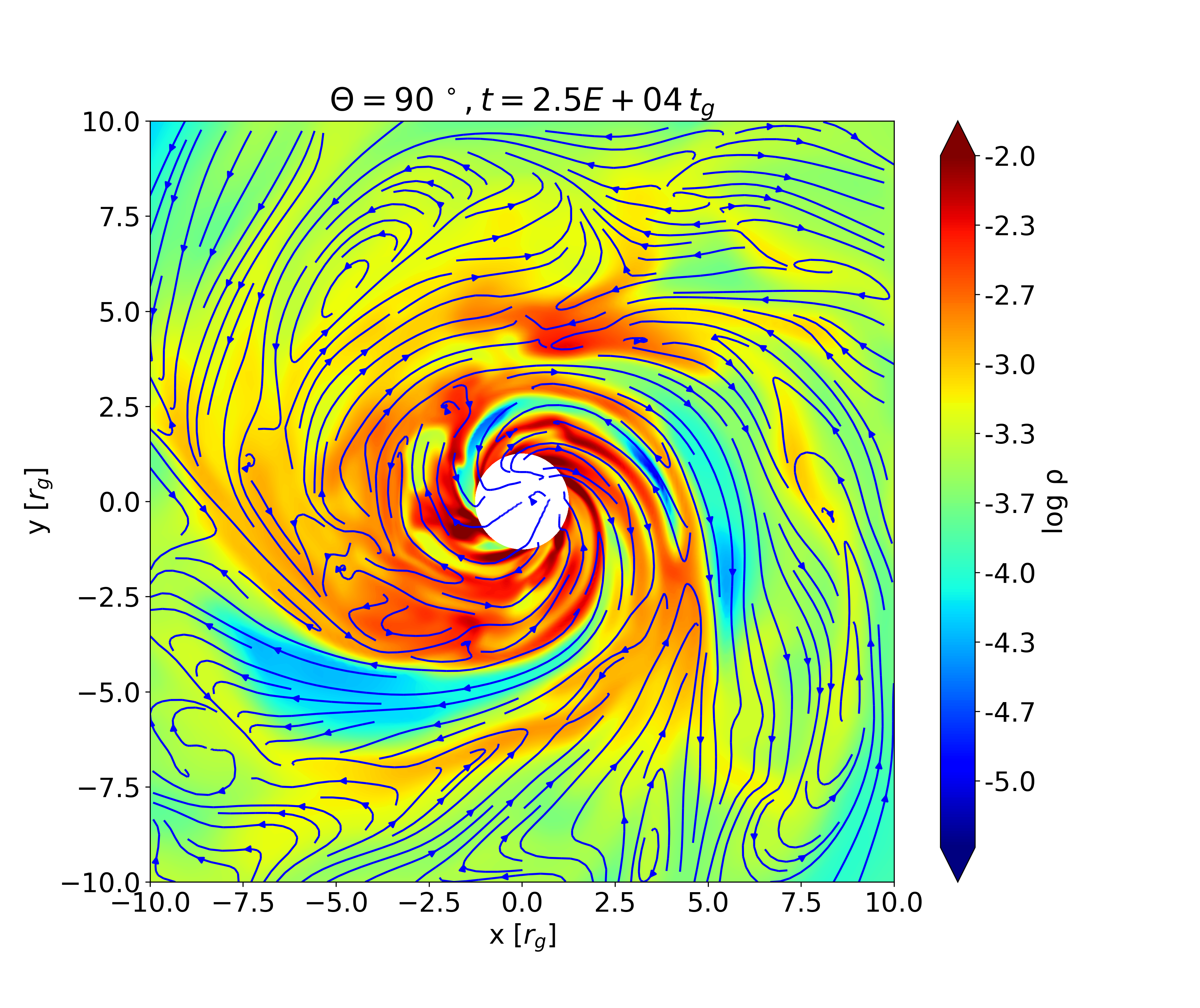}
    \caption{Simulation of GRB central engine in the magnetically arrested mode. Plot shows an equatorial cut of the 3D model, taken at evolved time of 25000 $t_{g}$. Color scale marks density of the gas, and the streamlines show magnetic field configuration}
    \label{fig:MAD_xy}
  \end{minipage}
  \quad
  \begin{minipage}{0.45\textwidth}
    \includegraphics[width=\textwidth]{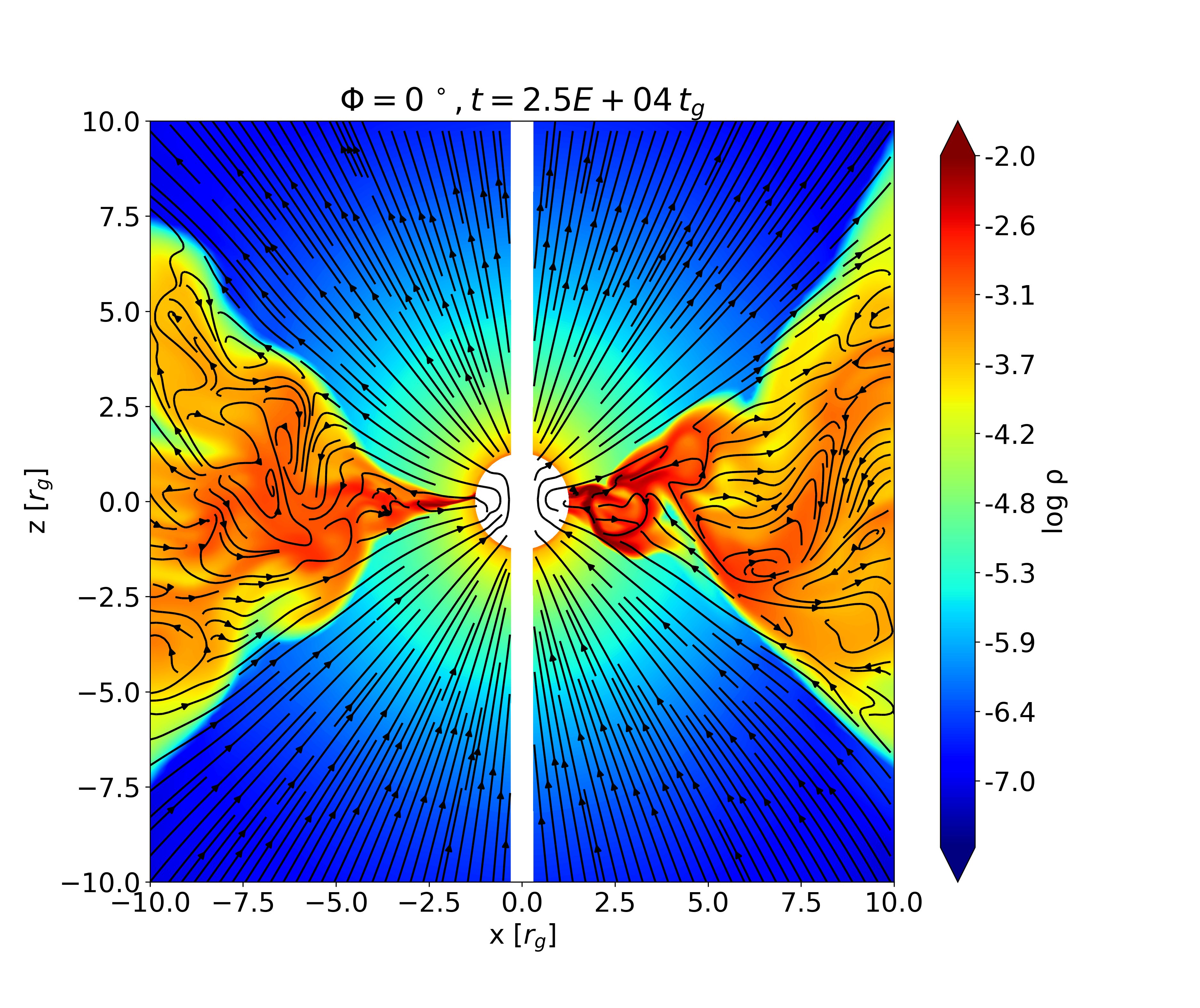}
    \caption{Simulation of the magnetically arrested accretion in the GRB engine. Plot shows a polar cut of the 3D model (the same time snapshot as in Fig. \ref{fig:MAD_xy}). Thin, asymmetric disk inside inner $\sim 5 r_{g}$ accretes through interchange instabilities. Open field lines form base of the jet}
    \label{fig:MAD_xz}
  \end{minipage}
\end{figure}

The magnetic field is dragged to the black hole horizon
during accretion, which leads to amplification of magnetic flux in the innermost region. While the magnetic
dynamo is sustained for a sufficiently long time, it can provide an efficient transport of
angular momentum via the magneto-rotational instability, on a global scale.
In the magnetically-arrested (MAD) state, a part of the disk is no longer in the MRI mode, and portions of matter can fall locally on dynamical timescale to the black hole. This is possible only due to the interchange (Rayleigh-Taylor type of) instability. Falling strips of material are pushed back by the magnetic field, but the magnetic flux is interchanged by mass flux at the disk equatorial plane. The accretion proceeds via low-order non-axisymmetric azimuthal modes of RTI, as can be seen in Figure \ref{fig:MAD_xy} showing the snapshot from our simulation.
This mode of accretion may explain the highly variable first couple hundred seconds of
the long GRB prompt emission \citep{LloydRonning2016}.

Figure \ref{fig:MAD_xz} presents the polar cut of the same 3D simulation state. The low-density polar regions are dominated by Poynting energy and highly magnetized. The power extracted via open magnetic field lines from the rotating black hole, scales in the
Blandford-Znajek process as $\dot{E}_{BZ} \propto \Phi^{2}_{BH} a^{2}/r_{g}^{2}$, and is naturally the highest for maximally spinning black holes.
The physical scaling of magnetic flux in gamma ray bursts and their progenitors has a characteristic order of $\Phi\sim 10^{18}$ G cm$^{2}$, with the horizon radius of $r_{g}\sim 4.5 \times 10^{5}$ cm. Hence, the power of jets extracted from rotating black hole in GRBs with this process can be as large as $10^{54}$ erg s$^{-1}$.

\section{Stellar collapse}

\subsection{Missing bright transient}

The maximal rotation of stars allowed for the collapsar to leave no bright transient and disappear without a trace, was compared with the simulated stellar evolution models from MESA code in \citet{2020ApJ...901L..24M}. It was shown that about 5\% of stars have the angular velocities below the critical value and will leave no accompanying GRB. Examples of such disappearing stars can be N6946-BH1
and PHL293B-LBV, two supergiant stars observed before outside the Milky Way, in metal-poor dwarf galaxies. In their emission spectra, the disappearance
of broad emission lines that were characteristic for an LBV star, suggest
that it could have collapsed to a black hole without a bright supernova
\citep{2020MNRAS.496.1902A}.

In our scenario, such collapsar that would not leave any accompanying bright transient, poses a limit on the stellar rotation at the onset of collapse.
Thus, we determined the critical angular velocity of a collapsing star, via numerical simulation. That initial simulation used a code HARM in its basic, publicly available version that
neglected either spin of the black hole and black hole mass change.
Therefore, the simulation setup enforced only a very short timescale to be covered.
We also neglected there a self-gravity of the accreting material.
All these factors can quantitatively change the results.

\subsection{Evolution of black hole mass and spin during collapse}

In Figures \ref{fig:mdot12}, \ref{fig:mbh12}, and \ref{fig:spin12}, we present
the time evolution of accretion rate onto black hole, mass of the black hole, and
its spin, during the collapse of massive star endowed with rotation.
The initial angular momentum per unit mass is given by $l=u_\phi=g_{\phi \nu}u^\nu$.
We include here a factor $C\sin^2\theta$ in the initial angular velocity profile such that it is scaled with the circular motion at the ISCO:
\begin{equation}
u^\phi=C \sin^2\theta (-g^{t\phi}\epsilon_{\rm isco}+g^{\phi\phi}l_{\rm isco}).
\end{equation}
The factor $\sin^2 \theta$ ensures that the angular momentum vanishes smoothly in the polar regions, as in \citep{2017MNRAS.472.4327S}, and $C$ is a parameter that we vary. The model with $C=0$ corresponds to Bondi spherical accretion.

In Figures \ref{fig:mdot12}-\ref{fig:spin12}, the two models represent accretion with critical, $C=1.0$, and
  supercritical, $C=2$ rotation. We neglect magnetic fields, and the initial black hole spin is zero. Mass of the star is 25 Solar masses.
  The accretion rate and black hole mass are expressed in $M_{\odot}$, while the spin parameter $a$ is dimensionless ($a=J/M^{2}$). Note that the time is expressed in geometric time units. When scaling with the initial black hole mass of 3 $M_{\odot}$, the time unit conversion will result in $t_{g}=G M_{BH}/c^{3}= 1.47\times 10^{-5}$ s, so the total simulation duration would be about 0.7 seconds.
  
\subsection{Effects of self-gravity in the collapsar}

\begin{figure}
  \centering
  \begin{minipage}{0.3\textwidth}
    \includegraphics[width=\textwidth]{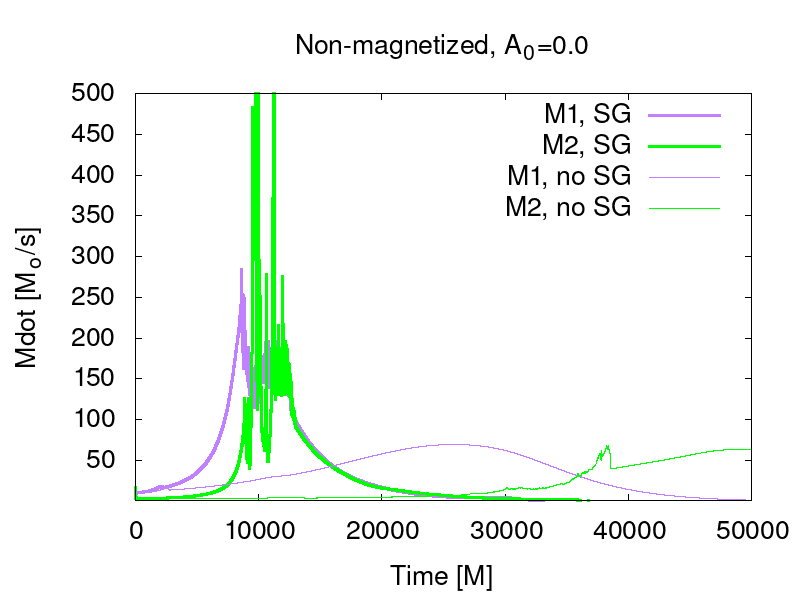}
    \caption{Accretion rate evolution. Models $M1$ and $M2$ have different normalisation of  angular momentum, $C1=1.0$ and $C2=2.$. }
    \label{fig:mdot12}
  \end{minipage}
  \quad
  \begin{minipage}{0.3\textwidth}
    \includegraphics[width=\textwidth]{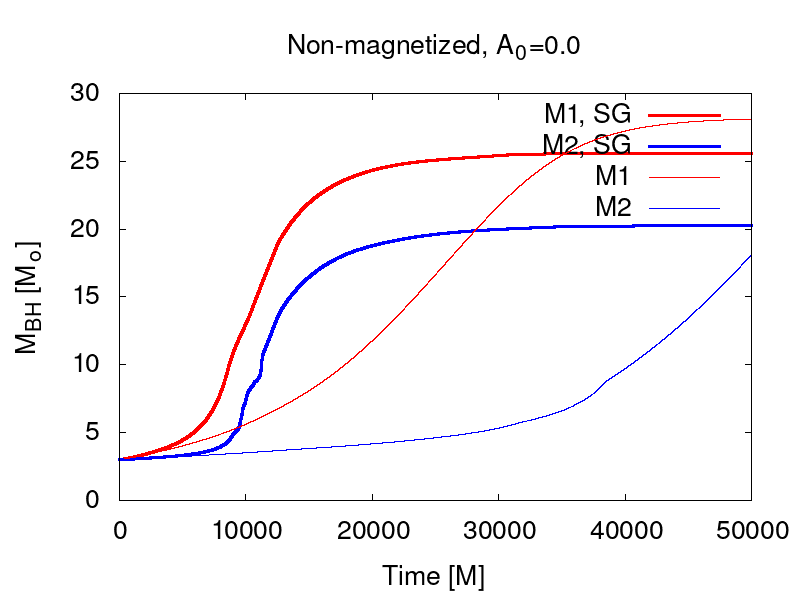}
    \caption{Black hole mass evolution. Initial mass of the black hole at the onset of collapse is 3 $M_{\odot}$, and mass of the star was 25 $M_{\odot}$}
    \label{fig:mbh12}
  \end{minipage}
  \quad
  \begin{minipage}{0.3\textwidth}
    \includegraphics[width=\textwidth]{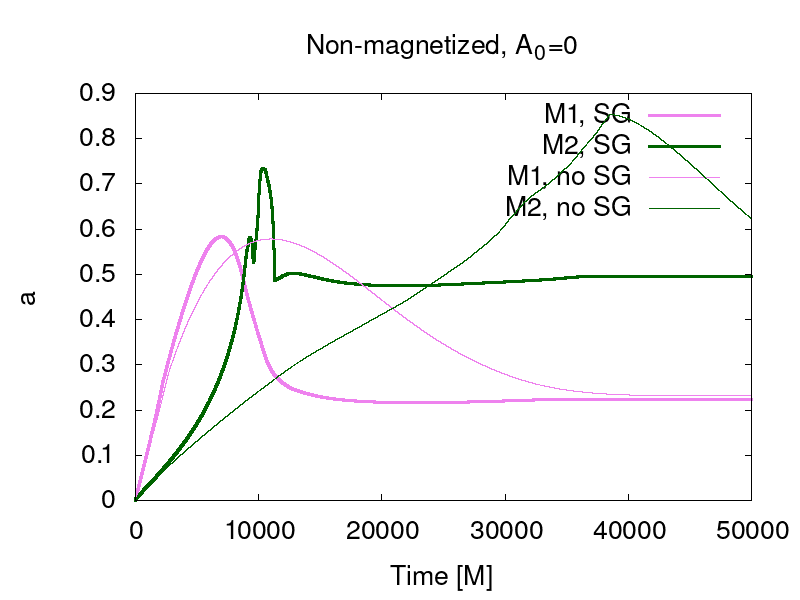}
    \caption{Black hole spin evolution. Initial spin of the black hole was $a=0$. Models $M1$ and $M2$ are the same as in Figs. 1 and 2}
    \label{fig:spin12}
  \end{minipage}
\end{figure}
In this section we presented
the preliminary results of the new implementation of the self-gravity perturbation in HARM-SELFG, 
developed
from  \cite{2021ApJ...912..132K}, where the Kerr metric coefficients were updated only due to the
increasing mass and spin of black hole, but the selfgravity perturbation was not accounted for.
The quantities plotted in Figures  \ref{fig:mdot12}- \ref{fig:spin12}, are showing the accretion rate through the BH horizon (left), increasing BH mass (middle), and evolving dimensionless spin (right, as a result of our GR-Hydrodynamical collapsar modeling.
Thick lines represent new models, with self-gravity of the accreting cloud. 
For comparison, with thin lines of the same colors, we also show the results from our older code version, without self-gravity (see \cite{2018ApJ...868...68J}).
We notice that the most dramatic effect on the accretion rate, and resulting black hole mass, is imposed by self-gravity at the initial phase of collapse. The mass of black hole grows much more quickly, due to self-gravity.
The black hole spin evolution is also affected, and obviously influenced by the amount of angular momentum in the stellar envelope. Hence, the final spin is the same for both SG and non-SG models. The maximum spin reached temporarily during collapse, is smaller in case of the SG model, but only for supercritical rotation of the envelope (models M2).

The self-gravity is accounted for with an approximate way, when the self-gravitating mass enclosed in the sphere of a radius $r$ at a given time $t$, is treated as a perturbation, additive to the black hole mass measured at this same time, at a point below the horizon radius. Hence, the gravitational influence of both $M_{BH}(t)$ and $\delta M(r,t)$ will affect material at all radii above this sphere.
The method is based on the idea of \cite{1972PhRvL..29.1114T} and for the current purpose, it was outlined in \cite{2020arXiv200507824P}.
The gravitating mass is a function of time and distance, and it reads:
\begin{equation}
M_{\rm grav}(r,t) = M_{BH}^{0} + \Delta M (t) + \delta M (r,t)
\end{equation}
where $M_{BH}^{0}$ is the initial mass of the black hole, $\Delta M (t)$ is the amount of mass-energy accreted through the black hole horizon until the given time, $t$, and $\delta M(r,t)$ is the perturbation of the black hole's gravitational potential, induced by the self-gravitating collapsar's mass, and integrated over the volume from the horizon up to the radius $r$.
A similar procedure is applied to the black hole spin, where the perturbation is added to the increasing black hole angular momentum, by integrating the angular momentum inside the sphere of the radius $r$, and adding it to the initial angular momentum of the black hole, and the angular momentum accreted through the horizon until the given time. Hence, we have:
\begin{equation}
J(r,t) = J_{BH}^{0} + \Delta J (t) + \delta J (r,t)
\end{equation}
while the dimensionless spin is defined as usual, $a=J/M$.
Our approximation is accounting for self-gravity, and the perturbative terms are incorporated in the Kerr metric evolution.

\section{Accretion under extreme conditions}

The plasma of accretion disk,
which constitutes the Gamma Ray Burst central engine, is extremely hot and dense $T>10^{9} K$, $\rho > 10^{10}$ g cm$^{-3}$.
Under such conditions, the microphysics of the gas is very important.
In simple models, the gas equation of state (EOS) is described by an adiabatic law, $p=(\gamma-1)u$, where $p$ denotes pressure, $u$ is internal energy, and $\gamma$ is an adiabatic index, of a value between $4/3$ and $5/3$.
Such EOS provides a closing equation for the system of magnetohydrodynamics equations that describe accretion flow evolution, and is frequently enough to
allow the code to capture an important physics. The phenomena such as the
jet launching via extraction of the black hole rotational energy, or the self-gravity of an accreting massive envelope, are well described by ideal MHD simulations in general relativity, and the gravitational force is the one that drives
the physical phenomena.

However, the nuclear forces come into the game, when we want to address other
complex observables, emerging on long timescales. These observables have recently brought a big excitement to the scientific community, with the discovery of a so-called kilonova signal, associated with the afterglow of a gamma ray burst.
It has been predicted \citep{LiPaczynski1998} that the radioactivities from dynamical ejecta after the binary neutron star merger can power an
electromagnetic signal. Its characteristics is similar to the supernova, with several orders of magnitude smaller energies, with the absolute magnitude below $M_{V}=-16$ at the peak, and rapidly fading, about $\sim 0.5$ mag per day.
These transients are also difficult to detect
because kilonovae are rare (their rate is estimated at below 1 per cent
of the core collapse supernova rate).
In addition, the subsequent accretion can provide contribution to the kilonova, if only the disk outflows are fast enough and not absorbed by precedent ejecta \citep{SiegelMetzger2017}. The heavy unstable isotopes are formed in the rapid neutron capture process (r-process), while the ejected material is very
neutron rich.
The blue and red kilonovae components associated with GW170817 have been attributed to  post-merger dynamical ejecta, and accretion disk winds \citep{Kilpatrick}. We model such winds, driven by magnetic fields and neutrino heating, and composed of a very dense, neutronized material, expanding homologously into the interstellar medium. 

\subsection{Nucleosynthesis in GRB accretion disk winds}

\begin{figure}
  \centering
  \begin{minipage}{0.45\textwidth}
    \includegraphics[width=\textwidth]{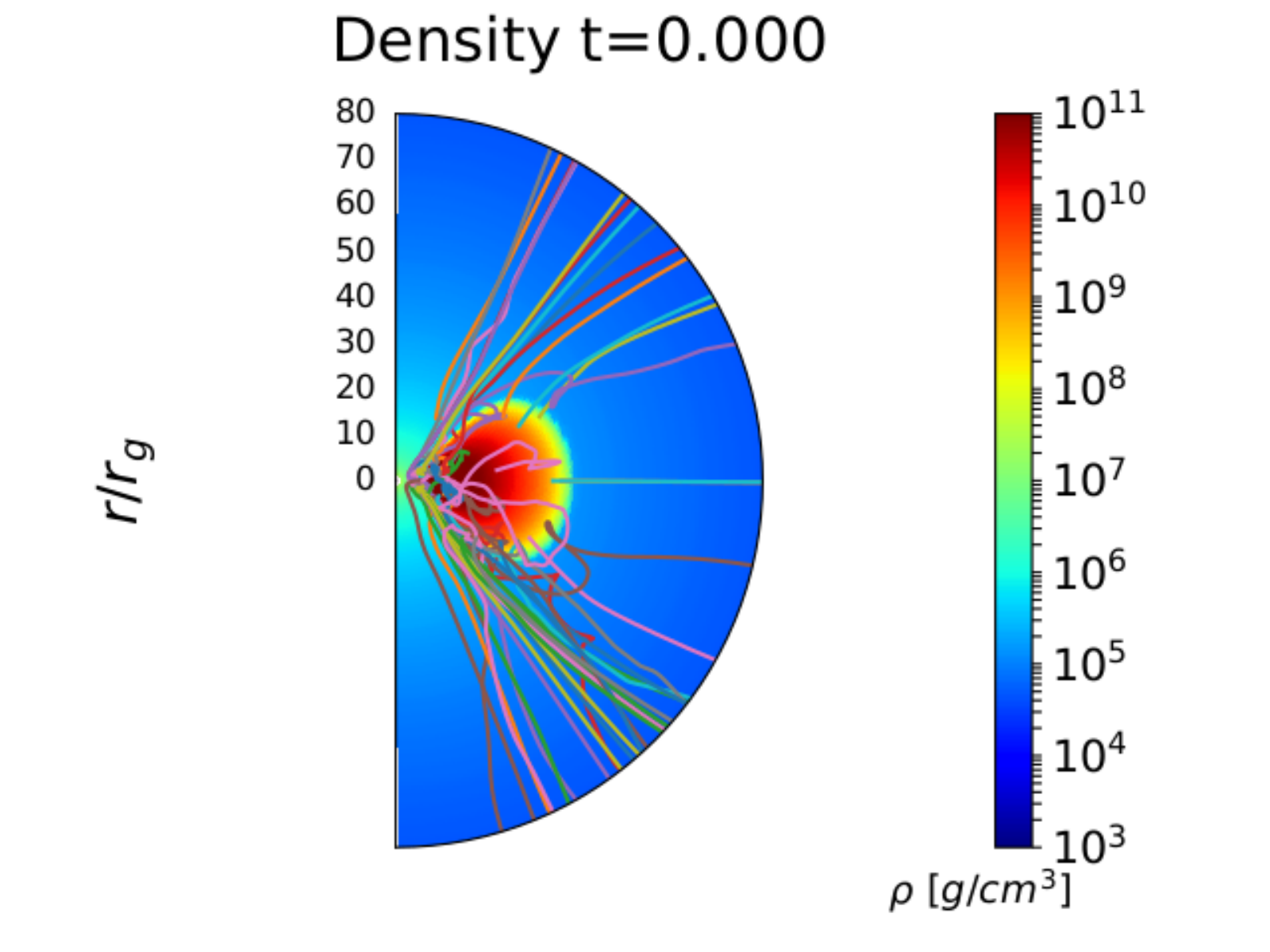}
    \caption{Tracing the accretion disk wind. Tracer particles are computed in time, until 20000 $t_{g}$, here overplotted on the disk density distribution at initial time}
    \label{fig:tracers}
  \end{minipage}
  \quad
  \begin{minipage}{0.45\textwidth}
    \includegraphics[width=\textwidth]{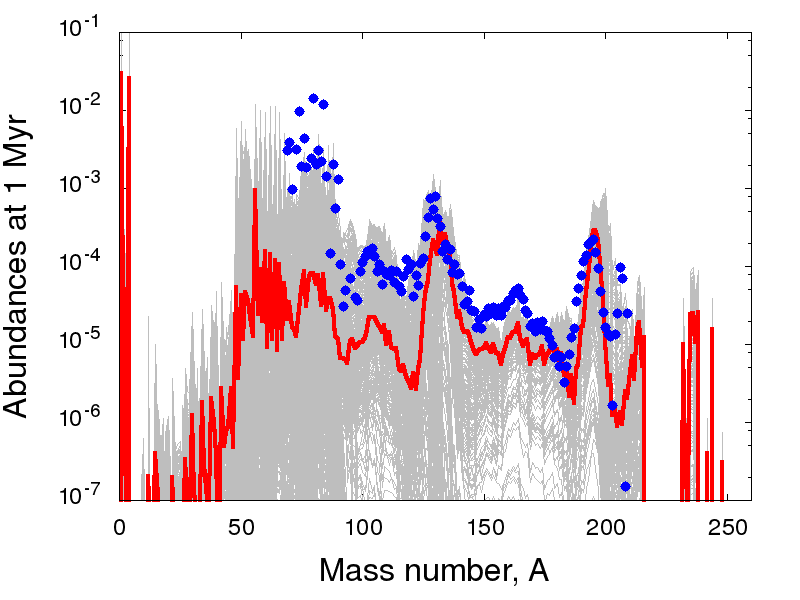}
    \caption{r-process nucleosynthesis in GRB accretion disk wind. Grey lines denote results on individual tracers; red line is the averaged model. Blue points are the Solar data}
    \label{fig:abund}
  \end{minipage}
\end{figure}

The two key elements are (i) tabulated equation of state, which substitutes the adiabatic law used in HARM code to model the disk accretion, and (ii) the usage of tracer particles, which enable to follow the nucleosynthesis on the seed heavy nuclei, expanding with the disk wind 
(see \citep{Janiuk2019}).
In the EOS, the contribution to the total pressure is given by the free nuclei, electron-positron pairs, Helium nuclei, radiation, and trapped neutrinos. The nuclei and pairs are relativistic and may have an arbitrary degeneracy level,
hence their chemical potentials are calculated. At every grid point, the
chemical composition is stored and updated over time, during the MHD simulation. 
The tracer particles are distributed uniformly inside the accretion disk, and
are initialized at the start of the simulation. We store separately the tracers, which leave our computational domain (located at about 1000 $r_{g}$) and use them as an input to the nuclear reaction network post-processing.
Sample tracers (not all, but a fraction of the outflowing particles) are visualized in Figure \ref{fig:tracers}

The nuclear reaction network is used to reproduce the abundance pattern of
r-process elements synthesized in these accretion disk outflows in GRB. The network contains more than a thousand isotopes in its database. The code takes into account the fission reactions and electron screening (see \cite{Lippuner2017} for details).
%{\bf more details?}
All three peaks of the Solar abundance pattern are reproduced in this way, where the first peak corresponds to the Iron group, and the second and third emerge from the Lanthanide- and Actinide-rich ejecta layers. The latter can be obtained only with the electron fraction on the order of $Y_{e}=0.1$ or less. As we have shown in Figure \ref{fig:abund}, a substantial mass fraction of the disk wind carries such a neutron-rich material.

The detailed composition pattern of the outflow depends however on the
engine parameters. In our case, it is the black hole spin and initial magnetisation of the disk, which scale the models.
We find, that both higher black hole spin and higher magnetisation drive faster and more massive winds (velocity up to $v/c\sim0.4-05$, mass up to $\sim 0.2$ of the initial disk mass). In this case, the distribution of electron fraction in the outflows is bimodal, and two peaks can be observed, around $Y_{e}=0.1$, and $Y_{e}=0.4$.
For the less magnetized disk around moderately spinning black hole, both velocity and mass of the wind ejecta is significantly smaller, but the $Y_{e}$ distribution is rather single-peaked, containing mostly material with electron fraction on the order of 0.1. We postulate therefore, that different types of engines,
will imprint their properties in the observed kilonova signals due to various composition patterns.

\section{Summary and Conclusions}

Gamma ray bursts are powered by narrow, ultra-relativistic jets of plasma
which can only be seen if the jet points towards the observer.
Such fast outflows are accelerated on the cost of power extracted from the rotating black
hole and are collimated by magnetic fields.
The common feature of both short and long GRB jets is that they
emerge from this accretion-powered engine. Further out, they have to
drill through a dense environment (merger ejecta or stellar envelope),
and generate the bright gamma-ray emission.
The observed high-energy emission is an ultimate probe of conditions inside the engine, and the progenitors physics, e.g. the formation of r-process elements in the post-merger ejecta,
or the explosion mechanism and structure of collapsing star. 
Because of extreme complexity of these systems, the semi-analytic approach which is still used in modeling of disk-jest systems such as Active Galactic Nuclei, or Black Hole X-ray Binaries, is no longer viable in the realistic GRB engine models.
On the other hand, non-linear nature of GR MHD equations, which dictate the jet physics, poses a challenge to any numerical simulations which aim for
large spatial and long temporal scales.
In practice, still the multi-scale picture is obtained by combining different type of simulation, and post-processing of the results. For instance, the expansion of jets through the post-merger wind in GW170817, was simulated on large scale with the Newtonian
code Flash, while the input wind models were derived by GR MHD and SPH simulations, to represent the magnetically- and neutrino-driven winds, respectively \citep{Murguia2021}.

To date, despite many attempts done by various groups, there is no single GR simulation which would be able to explain the collapsar evolution, starting from a pre-supernova phase, through the black hole formation and spin-up, until the jet break through the ejected envelope. Long-term simulations with
the most advanced input microphysics are still missing either, to fully understand the development of relativistic jet in BNS merger and the origin of kilonova.
Our proceeding summarized a few attempts to build a reliable simulation of black hole accretion disk and jet engine of either long or short GRB, where the spacetime
evolution, and detailed microphysics, are carefully taken into account. Still, there is a lot to be done, and models must be confronted with the future multi-messenger observations of collapsing or coalescing stars.

\acknowledgements{This research was supported in part by the grants 2016/23/B/ST9/03114 and 2019/35/B/ST9/04000 from Polish National Science Center. This research was carried out with the support of the Interdisciplinary Center for Mathematical and Computational Modeling at the University of Warsaw (ICM UW) under grant no g85-986.
  This research was also supported in part by PLGrid Infrastructure.
}

\bibliographystyle{ptapap}
\bibliography{janiuk}

\end{document}